\documentclass[twocolumn,twoside,slac_two]{revtex4}
\usepackage{graphicx}
\usepackage{fancyhdr}
 \bibpunct{[}{]}{,}{n}{}{,}%
 
\usepackage[]{natbib}
\def\ltsima{$\; \buildrel < \over \sim \;$}
\def\simlt{\lower.5ex\hbox{\ltsima}} 
\def\gtsima{$\; \buildrel $>$ \over \sim \;$}
\def\simgt{\lower.5ex\hbox{\gtsima}} 

\def\phflux{photons cm$^{-2}$ s$^{-1}$}

\def\grayh{$\gamma$-ray}
\def\grays{$\gamma$-rays}

\def\dtr{\hbox{$\Delta$$t_{\rm r}$}}

\def\p0{$\pi^{\rm 0}$}

\def\r95{$r_{\rm 95}$}

\def\Fermi{\textit{Fermi}}

\def\bi{B0218+357}

\begin{document}

\title{\Fermi-LAT Detection of a Hard Spectrum
Flare from the Gravitationally
Lensed Blazar \bi}

%

\author{S. Buson}
\affiliation{Universit\`a di Padova, Dipartimento di Fisica e Astronomia ``G. Galilei'', I-35131 Padova, Italy\\
        INFN, Sezione di Padova, I-35131 Padova, Italy}
\author{C. C. Cheung}
\affiliation{Space Science Division, Naval Research Laboratory, Washington, DC 20375-5352, USA}
\author{S. Larsson}
\affiliation{Department of Physics, Stockholm University, AlbaNova, SE-106 91 Stockholm, Sweden\\
The Oskar Klein Centre for Cosmoparticle Physics, AlbaNova, SE-106 91 Stockholm, Sweden\\
Department of Astronomy, Stockholm University, SE-106 91 Stockholm, Sweden
}
\author{J. D. Scargle}
\affiliation{Space Sciences Division, NASA Ames Research Center, Moffett Field, CA 94035-1000, USA}
\author{on behalf of the \Fermi-LAT Collaboration}
 \affiliation{ }
 
\begin{abstract} \noindent
The \Fermi-LAT has observed new \grayh\ flares from the blazar B0218+357 during July 2014. While no significant change in the \grayh\ spectrum has been previously observed through the flaring phase in late-2012, during this recent high activity the source displayed an exceptionally hard spectrum. The latter led to the detection of very high energy (VHE, $E > 100$ GeV) \grays\ from B0218+357 by the MAGIC telescopes, establishing this source as the most distant TeV emitter known to date. In addition to the detection of VHE emission, this blazar is of particular interest since it is known to be a double-image gravitationally lensed system with a lens delay of $11.46 \pm 0.16$ days measured
in \grays . 
We present the \Fermi-LAT study of the July 2014 flares and discuss them in the context of previous measurements.
\end{abstract}

\maketitle
\thispagestyle{fancy}

\section{Introduction}
\bi\ is a blazar at $z=0.944\pm0.002$ \cite{cohen03} lensed by a galaxy at 
redshift $z=0.6847$ \cite{browne93}. The system appears as a double image in radio separated by 335 milli-arcseconds with a brighter  western A and fainter eastern B images  and an Einstein ring \cite{odea92,patnaik93}. The delay between the two images has been measured in radio as $\dtr=10.5\pm0.2$ \cite{biggs99} and $10.1\pm0.8$ days \cite{cohen00}. 
Further timing analysis of the radio light curve from  \cite{cohen00} using two independent methods resulted in  two other possible delay values, $\dtr=9.9^{+4.0}_{-0.9}$ or $11.8\pm2.3$ days \cite{eulaers11}.
\bi\  was detected for the first time in \grays\  by the \Fermi\ Large Area Telescope \cite[LAT;][]{atwood09}.  Over the initial 4-years of LAT observations, \bi\ displayed a \grayh\ flux of $F_{E > 100 \mathrm{MeV}}=(2.4\pm0.9)\times10^{-8}$ \phflux\ and a photon index of $\Gamma=2.28\pm0.03$ \cite[3FGL~J0221.0+3556; third \Fermi-LAT Catalog, 3FGL,][]{3fgl}.

\section{Gamma-ray flares from \bi}
During the first six years of \Fermi-LAT operations \bi\ has displayed two major phases of enhanced activity. The first was a long-lasting one that developed between 2012 June and 2013 March. It was characterized by a series of prominent \grayh\ flares that persisted over four months. A detailed study of this period led to the measurements of the first robust \grayh\ lensing delay.
Applying an auto-correlation function (ACF) to the evenly sampled light curve focused on the 2012-2013 enhanced \grayh\ activity, a single dominant correlation peak was apparent. Its significance was evaluated to be 9$\sigma$. The delay in the \grayh\ data was inferred to be $11.46 \pm 0.16$ days (1$\sigma$).
With the same dataset it was possible to estimate the flux ratios (A/B), which were found to be consistent with unity, and similarly, infer a magnification ratio of about unity \cite{cheung14}.
Importantly, during this period no indication for spectral variability was found. The \grayh\ photon index $\Gamma= 2.30 \pm 0.03$ was consistent with the value derived before the start of the flaring; also, cf. the 3FGL value.

\bi\ underwent a second enhanced phase during 2014 July. The flares 
were between 2014 July 11 and 27 (MJD 56849--56865) remarkably characterized by a distinctive hard spectrum \cite{bus14Atel}.
Based on a preliminary analysis, on 2014 July 13 and 14 (MJD 56851 and MJD 56852) the source was observed with respective daily averaged fluxes ($E >$ 100~MeV) of $(6.5 \pm 1.4) \times 10^{-7}$ \phflux\ with  $\Gamma = 1.4 \pm 0.1$ and $(6.7 \pm 1.5) \times 10^{-7}$ \phflux\ with $\Gamma = 1.6 \pm 0.1$.
Because the gamma-ray delay was previously measured, LAT Target of Opportunity observations were triggered to trace the expected delayed emission, about 11 days after. Noteworthy, emission at VHE was subsequently reported by the MAGIC telescopes at the time of the expected delayed \grayh\ flare 
\cite{mirzoyan14,mazin15}.
%
\begin{figure*}
\includegraphics[width=0.8\textwidth]{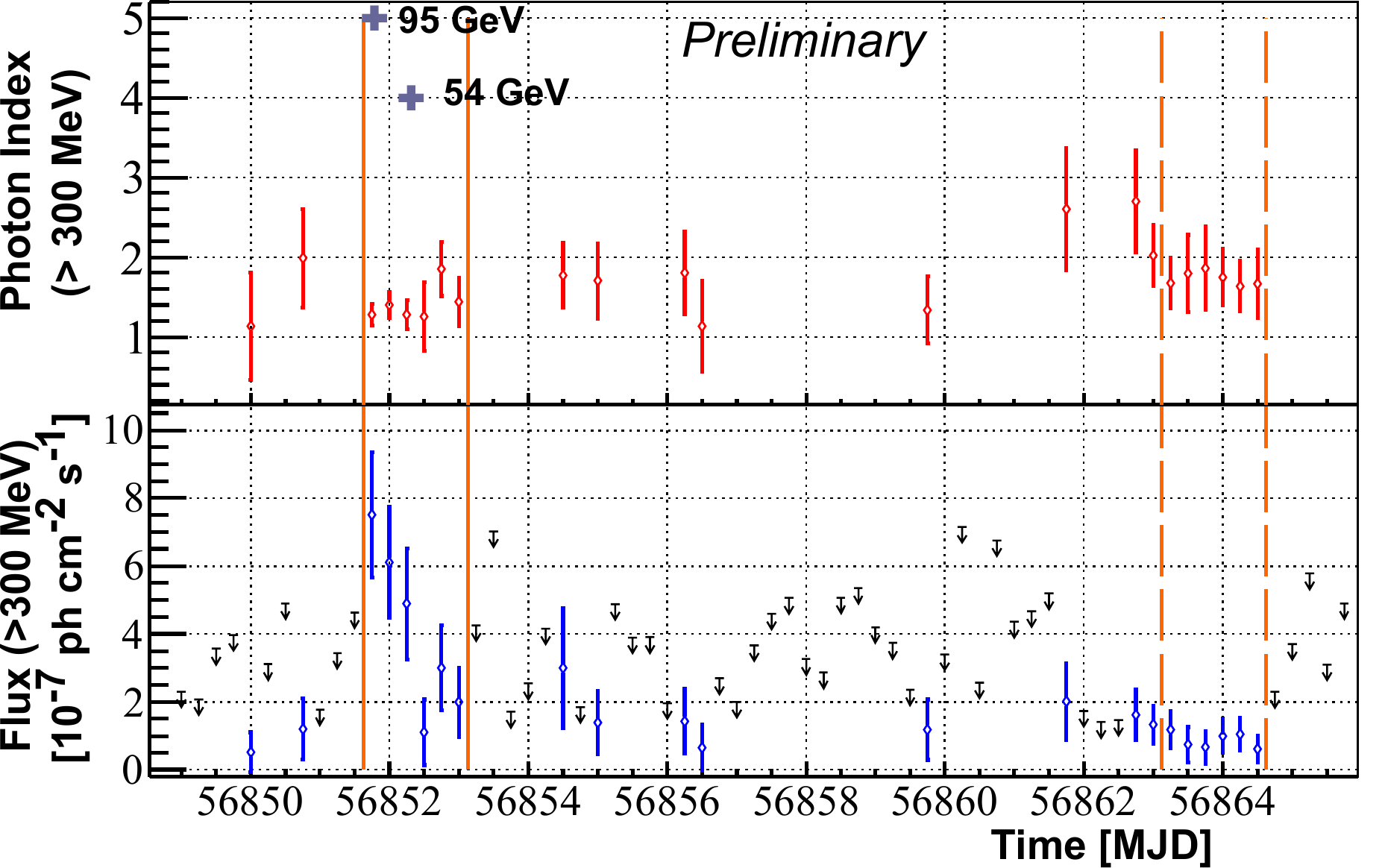}%
\caption{
\Fermi-LAT light curve computed in 6-hr bins at energies $E > 300$ MeV (lower panel) and the corresponding photon indices (upper panel). 
Vertical lines indicate the intervals of the flare (solid) and corresponding delayed emission (dashed).
The two crosses indicate the arrival time of the two $>$50 GeV events (see Section \ref{sec:ana} for more details). }\label{fig:lc}
\end{figure*}
In the following we present the preliminary analysis of the 2014 LAT dataset.

\section{Hard spectral phase during 2014}

\subsection{\Fermi-LAT analysis}\label{sec:ana}
\Fermi-LAT data have been analyzed as described in \cite{cheung14} but restricted to the energy range 300 MeV -- 300 GeV to better characterize the hard-spectrum flaring behavior reported by \cite{bus14Atel} (ATel \#6313). The source was modeled with a single power law with photon index free to vary. In the analysis, 95\% confidence level upper limits were computed when the test statistic\footnote{The TS corresponds roughly to the square of the significance assuming one degree of freedom.} (TS) \cite{mattox96} for the source was TS $<$ 8.
A preliminary study of data between MJD 56849--56865 confirms the detection of \bi . During this interval the photon index is found to be on average exceptionally smaller (below 2) with respect to the value of 2.3 reported in previous LAT studies \cite{cheung14}.
Noteworthy, two high energy photons are positionally consistent with the source: a 95 GeV and a 54 GeV event, detected at $\sim$MJD 56852, i.e. at the time of the initial flare.

\subsection{Light Curve}
Figure \ref{fig:lc} presents the 6-hr light curve ($>$300 MeV)
and the corresponding photon index versus time variations covering the time interval MJD 56849--56865.
A first flaring phase, corresponding to flare of image A, is well defined by a fast flux increase accompanied by an exceptional hard spectrum and lasted 1.5 days. This phase is identified by six consecutive 6-hr source detections, as denoted in Figure \ref{fig:lc} by the two continuous vertical lines between MJD 56851.625--56853.125.
Immediately after this short flare the source returned to quiescence, displaying only sporadic activity  for the subsequent $\sim$11 days.

Then, at the time of the expected delayed emission, the source is observed again as denoted by the series of bins between the two dashed vertical lines, at MJD 56863.125--56864.625. 
These detections happen $\sim$11.5 days later. The six consecutive 6-hr  intervals with TS $>$ 8 are distinguished again by a small photon index ($\Gamma<2$).
This second flaring phase corresponds to the flare of image B, based on the hard-spectrum signature.
Formally, the ACF computed on the LAT light curve of 1-day binning between MJD 56849-56865 shows a peak around 11 days (not shown). Though its statistical significance has not yet been estimated it is notable that the peak is broadly consistent with the previous \grayh\ delay of $11.46 \pm 0.16$ days.

We compare the flux measured during the first 1.5-day flaring episode, beginning at MJD 56851.625, with the observation shifted by 11.5 days. We derive the resultant average flux ratio A/B $\sim 4$.
 \begin{figure}[]
\includegraphics[width=0.4\textwidth]{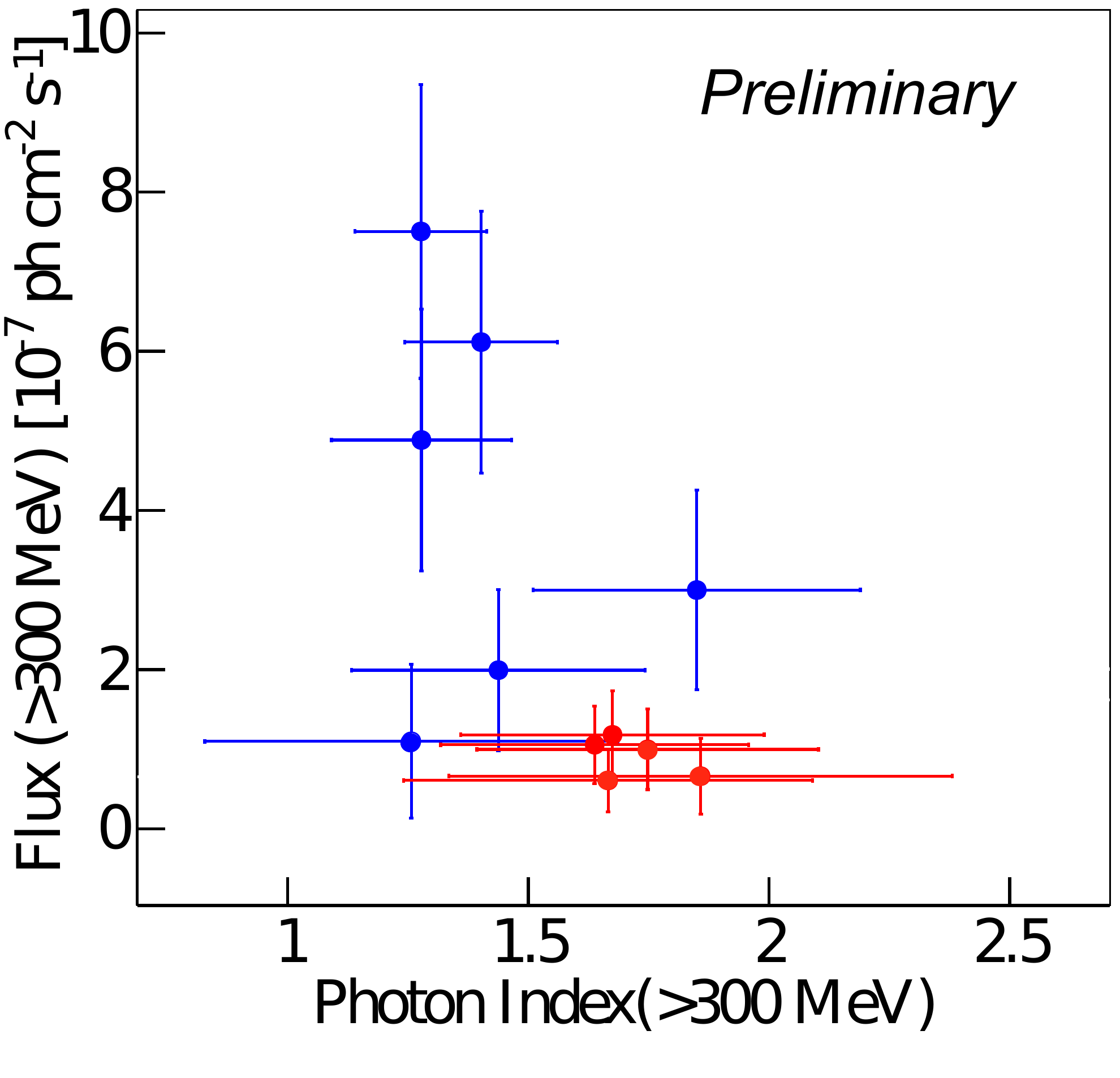}%
\caption{Scatter plot of the 6-hr binned fluxes and photon indices $E > 300$ MeV for 
the 1.5-day interval beginning at MJD 56851.625 (blue) and beginning at MJD 56863.125 (red), the same intervals reported in Figure \ref{fig:lc} between solid and vertical lines, respectively.
}
\label{fig:scatter}
\end{figure}

\section{Summary}
Previous analysis of LAT data for the 2012 flares of the gravitationally lensed blazar \bi\ found a \grayh\ delay measurement of $11.46 \pm 0.16$ days and a flux ratio estimate between images (A/B) consistent with $\sim$1. 
From a preliminary analysis of the 2014 July LAT data, we found:
\begin{itemize}
  \item Flares are characterized by peculiar hard spectra, with  photon index $\Gamma < 2$;  
   \item Based on the hard spectral signature of the flaring emission, the delay is compatible with the previous \grayh\   delay measurement of 11.5 days;
   \item The average flux ratio (A/B) over this flaring episode is $\sim$ 4: while this factor is different with respect to the previous \grayh\ measurement, it is interestingly closer to the radio observations ($\sim$3 or 4) \cite{biggs99,cohen00};
   \item A 95 GeV and a 54 GeV photon were detected by the LAT during the first flare (A image) but were not observed during the second flare of image B. Nevertheless, the MAGIC detection ensures that in the second phase the source emission extends up to $\sim$100--200 GeV.
\end{itemize}

\bigskip 
\begin{acknowledgments}
Work by C.C.C. was supported at NRL by NASA DPR S-15633-Y and Guest Investigator program 13-FERMI13-0009. S.L. acknowledges support by a grant from the Royal Swedish Academy Crafoord Foundation. 
The \textit{Fermi}-LAT Collaboration acknowledges support for LAT development, operation and data analysis from NASA and DOE (United States), CEA/Irfu and IN2P3/CNRS (France), ASI and INFN (Italy), MEXT, KEK, and JAXA (Japan), and the K.A.~Wallenberg Foundation, the Swedish Research Council and the National Space Board (Sweden). Science analysis support in the operations phase from INAF (Italy) and CNES (France) is also gratefully acknowledged.
\end{acknowledgments}
 \bibliographystyle{../Scineghe2014/aa}
\bibliography{/Users/buson/Presentazioni/Proceeding/full_biblio} 

\end{document}